**Title:** Dynamically Adjusting Case Reporting Policy to Maximize Privacy and Public Health Utility in the Face of a Pandemic

**Authors:** J. Thomas Brown, BS[1]; Chao Yan, PhD[2]; Weiyi Xia, PhD[1]; Zhijun Yin, PhD[1,2]; Zhiyu Wan, PhD[1,2]; Aris Gkoulalas-Divanis, PhD[3]; Murat Kantarcioglu, PhD[4]; Bradley A. Malin, PhD[1,2,5]

**Affiliations:**
[1]Department of Biomedical Informatics, Vanderbilt University Medical Center, Nashville, TN, USA
[2]Department of Electrical Engineering and Computer Science, Vanderbilt University, Nashville, TN, USA
[3]IBM Watson Health Cambridge, MA, USA
[4]Department of Computer Science, University of Texas at Dallas, Dallas, TX, USA
[5]Department of Biostatistics, Vanderbilt University Medical Center, Nashville, TN, USA

**Corresponding Author:**
J. Thomas Brown
Address: 2525 West End Ave, Suite 1475, Nashville, TN 37203
Email: james.t.brown@vanderbilt.edu
Telephone: (801) 608-2229

## ABSTRACT

**Objective:** Supporting public health research and the public's situational awareness during a pandemic requires continuous dissemination of infectious disease surveillance data. Legislation, such as the Health Insurance Portability and Accountability Act of 1996 (HIPAA) and recent state-level regulations, permits sharing de-identified person-level data; however, current de-identification approaches are limited. Namely, they are inefficient, relying on retrospective disclosure risk assessments, and do not flex with changes in infection rates or population demographics over time. In this paper, we introduce a framework to dynamically adapt de-identification for near-real time sharing of person-level surveillance data.

**Materials and Methods:** The framework leverages a simulation mechanism, capable of application at any geographic level, to forecast the re-identification risk of sharing the data under a wide range of generalization policies. The estimates inform weekly, prospective policy selection to maintain the proportion of records corresponding to a group size less than 11 (PK11) at or below 0.1. Fixing the policy at the start of each week facilitates timely dataset updates and supports sharing granular date information. We use August 2020 through October 2021 case data from Johns Hopkins University and the Centers for Disease Control and Prevention to demonstrate the framework's effectiveness in maintaining the PK11 threshold of 0.01.

**Results:** When sharing COVID-19 county-level case data across all US counties, the framework's approach meets the threshold for 96.2% of daily data releases, while a policy based on current de-identification techniques meets the threshold for 32.3%.

**Conclusion:** Periodically adapting the data publication policies preserves privacy while enhancing public health utility through timely updates and sharing epidemiologically critical features.

## INTRODUCTION

The novel coronavirus 2019 (COVID-19) pandemic has put a spotlight on infectious disease surveillance systems[1] and the importance of making such information widely accessible[2]. Sharing surveillance data in a timely manner can support a wide variety of public health research endeavors (e.g., from modeling disease transmissibility to simulating interventions[3–6]) and provide the public with situational awareness of outbreaks[4,7,8]. In recognition of such benefits, over the past year and a half, various organizations have worked to broaden access to large epidemiological datasets. Recent instantiations of COVID-19 initiatives include the National COVID Cohort Collaborative (N3C) of the U.S. National Institutes of Health[9], the Datavant COVID-19 Research Database[10], the Centers for Disease Control and Prevention's (CDC) COVID-19 Case Surveillance datasets[11–13], and the Global.health data science initiative[14], among others.

While advances in surveillance have spurred rapid growth in the volume and diversity of epidemiological resources, public data sharing on a wide scale remains limited[15]. This is due to numerous social and political factors, but it is evident that privacy is a core driving factor. In the United States, for instance, infectious disease data is captured by a variety of organizations, such as public health authorities, hospitals, and pharmacies. In regard to public data dissemination, such organizations may be subject to the Health Insurance Portability and Accountability Act of 1996 (HIPAA) and related laws and policies. Under HIPAA, an organization is permitted to publicly share patient-level data only when it is de-identified, that is, when "there is no reasonable basis to believe that the information can be used to identify an individual."[16] Even when organizations are not covered by HIPAA, they may be permitted to share data in a de-identified form as well. For example, the California Consumer Protection Act, the Virginia Consumer Data Protection Act (VCDPA), and the Colorado Privacy Act provide exemptions to de-identified data sharing[17–19]. However, transforming data into a de-identified form is a non-trivial endeavor. Numerous demonstration attacks have shown that, with the right background knowledge, a data recipient can leverage residual information in the records to re-identify the individuals to whom the data corresponds[20–25]. Concerns over such intrusions to anonymity have discouraged organizations from sharing data[26,27], which raises the importance of the question: How can organizations best comply with regulatory requirements while making surveillance data publicly available?

Under HIPAA, de-identification can be satisfied through two alternative implementations. The first is Safe Harbor, which requires the suppression of eighteen direct (e.g., patient name) and quasi-identifying features (e.g., geocodes with populations smaller than 20,000 residents). However, Safe Harbor requires hiding epidemiologically critical factors, such as reducing the granularity of dates of events to their year, which renders such a policy useless for characterizing infectious disease transmission. The alternative is Expert Determination, which indicates data is de-identified when "the risk is very small that the information could be used to identify an individual who is a subject of the information."[28] Various methods for risk assessment have been developed, including those previously developed for surveillance data[29], but provide limited guidance on adapting policies to the needs of the moment. Rather, they are retrospective in nature in that they assume data have already been collected and are ready for dissemination. Moreover, most methods further assume the number of records in the dataset remains fixed[30]. These assumptions differ from the requirements of case reporting while in the face

---

of a pandemic. Waiting to publish the data will hinder the ability to characterize the current state and evolution of an outbreak[1,2,31,32]. The infection rate must also be considered in the de-identification approach as it directly and dynamically influences the number of records in the dataset. Furthermore, several factors affect the privacy risk, including the demographics of the people infected[20,22] and the geolocations to which the pandemic spreads[33,34]. These requirements motivate the need for methods that forecast surveillance data.

In this paper, we introduce an approach to adaptively generate policies to publicly share de-identified patient-level epidemiological data. The framework simulates disease cases to estimate the longitudinal privacy risk of sharing infected individuals' quasi-identifier information at different levels of granularity in the absence of actual patient data. Periodically adjusting the policy allows the data sharer to adapt data granularity according to the influx of new patient records, while simultaneously allowing periods of consistent quasi-identifier representation. We specifically apply the framework to illustrate how policies could be developed to share COVID-19 patient health information and compare such policies to a more traditional de-identification approach relying on retrospective risk assessment. Furthermore, to be consistent with the CDC's current practice of using generalization and suppression for privacy[13], we use the framework to explore a wide range of data generalization policies.

It should be recognized the framework applies to any type of epidemiological disease spread, adjusts for the demographic diversity of individual US counties, and relies on public data sources. The framework can also be reused to address emerging data sharing needs, such as for vaccine registries[35,36]. Dynamically adapting data sharing policies holds the potential to consistently share more data with the public in a timely and privacy-preserving manner, fueling our data-driven response to infectious disease[8].

**MATERIALS AND METHODS**

Due to the challenge of predicting exactly who will be infected, prospectively fixing a data sharing policy requires probabilistic risk assessment. Our framework provides longitudinal privacy risk estimates for a data generalization policy within a specified geographic region. Given the appropriate population statistics, the framework can utilize any geographic level of detail (e.g., state, county, or ZIP code). In this research, we apply the framework to simulate disease spread on a county level to match the format of the COVID-19 surveillance data made accessible by the CDC[11,12]. In this section, we summarize the framework's features and its application to contextualize the results. Specific technical details are provided in the Supplementary Information.

**Privacy risk estimation framework**
Figure 1 summarizes the framework. In the first step, we select a data generalization policy, which defines the generalization of each quasi-identifying feature considered. In this paper, we consider basic demographic features and the date of diagnosis as quasi-identifiers, as they are typical features organizations have been requested to share (Table 1). The second step generates the county-level population across the quasi-identifying features per the selected policy. We use population count data from the U.S. Census Bureau to calculate the number of people in the county that fall into each demographic group[37], where each group is defined by a unique combination of quasi-identifier values, excluding date of diagnosis.

---

*The complete version of this paper is available, open-access, in the Journal of the American Medical Informatics Association: https://doi.org/10.1093/jamia/ocac011

**Table 1.** The quasi-identifiers considered in this study. The middle column describes the generalization strategy for each quasi-identifier. The third column provides an example generalization for each quasi-identifier. In the case of sex and ethnicity, the information is either included or null. AIAN = American Indian/ Alaskan Native, PI = Pacific Islander. *These values cannot be generalized since we simulate on a county level. †This definition of a week is consistent with the one used by the CDC's COVID-19 case forecasts[38].

| Field | Generalization Strategy | Generalization Example |
| --- | --- | --- |
| State of residence | None* | NA |
| County of residence | None* | NA |
| Date of diagnosis | Combine into week ranges (Sunday-Saturday†) | 01/05/21 → 01/03/21 - 01/09/21 |
| Year of birth | Convert to age ranges | 1980 → 40-45 years old |
| Sex | Nullify value | Female → null, Male → null |
| Race | Combine race groups | AIAN → AIAN or PI, PI → AIAN or PI |
| Ethnicity | Nullify value | Hispanic-Latino → null, Non-Hispanic → null |

The third step applies a Monte Carlo simulation (represented by the black box in Figure 1) to generate synthetic patient datasets using the county-level population distribution and a time series of new disease case counts. The time series' periodicity defines the frequency at which the updated dataset is released (e.g., every day or every week). To simulate the COVID-19 pandemic, we input time series derived from the Johns Hopkins COVID-19 tracking data[39]. The simulation algorithm (details of which are in the Supplementary Information) initially assumes that the no one in the county is infected. Then, for each time point, we randomly sample the number of disease cases (without replacement) from the uninfected population to form the newly reported patient dataset. The framework assumes individuals are not re-infected (for simplicity, considering a potentially negligible COVID-19 reinfection rate[40]) and assumes equal weighting across all individuals when sampling (to model the general uncertainty of disease spread, particularly in pandemics[41]).

The algorithm computes the re-identification risk on the patient set at each time point, according to a specified risk measure. There are various methods for measuring privacy risk[30]. In this work, we measure risk as the proportion of individuals in the dataset that fall into a group of size less than $k$, where each group is defined by a unique set of quasi-identifier values[42,43]. We refer to this measure as the *PK risk* and evaluate it given a set of $k$ values (as defined below) consistent with the standard thresholds used by public

---

health authorities[44–48]. The PK risk assumes a data recipient knows 1) an individual is a member of the dataset, 2) the individual's name and quasi-identifying information, and 3) the individual's relative date of diagnosis for the disease of interest. In this scenario, the data recipient attempts re-identification to learn the target individual's sensitive information from additional features included in the dataset (e.g., comorbidities[49,50]). The more unique the record's representation, the more likely the data recipient can re-identify the individual[20,22]. In this research, we focus on this risk measure to follow the CDC's application of $k$-anonymization[51]. The PK risk effectively measures the proportion of records that fail to achieve $k$-anonymity.

In practice, obtaining such patient information is difficult[23,52]. Thus, evaluating the PK risk provides an upper bound of re-identification risk for the dataset. To demonstrate the approach's flexibility and to offer a different perspective on privacy risk, we further analyze the amortized re-identification risk[53] in the Supplementary Information. The amortized re-identification risk relaxes assumptions (1) and (3) and considers the scenario in which the data recipient is motivated to re-identify as many patients as possible to learn who has the infectious disease of interest.

We highlight that, when applying the PK risk measure, we assume the attacker knows the diagnosis occurred within a lagging period of time (e.g., within one, three, or five days prior to the documented date). We allow this flexible assumption as it is unlikely a data recipient knows the targeted individual's exact diagnosis date[54], particularly when the time from a diagnostic test to case report extends beyond one day. The group corresponding to an individual contains all patients in the simulated patient set that match the individual on the demographic features, with a diagnosis date falling within the lagging period.

The final step of the framework uses the privacy risk distributions to estimate when the policy meets a privacy risk threshold. Computing the longitudinal privacy risk estimates under several data sharing policies for the same county identifies which policies likely meet the threshold at each point in the time series. The data sharer can then choose which policy to apply according to information priorities (e.g., prioritizing age granularity over sex granularity).

**Dynamic policy search**

To dynamically adapt policies according to an expected infection rate, we identify policies that are likely to satisfy a specific PK risk threshold at varying volumes of new case records. For this policy search, we choose a $k$ of 11, which is as a typical group size incorporated into guidance issued at the state[45–48] and federal[44] level. It is also the group size applied to CDC's COVID-19 Public Use Data with Geography[11]. We henceforth refer to the PK risk when $k$ equal to 11 as the PK11 risk. In this paper, we search for policies that meet a PK11 threshold of 0.01; i.e., the percentage of records falling into a demographic group of size 10 or smaller should be less than or equal to 1%. Similar investigations for $k$ of 5 and 20 (other common group size thresholds) are provided in the Supplementary Information.

The search uses the privacy risk estimation framework to evaluate 96 alternative data sharing policies for each U.S. county (with available census tract information) across a range of case count values. The policies include six potential generalizations of age, four generalizations of race, two generalizations of sex, and two generalizations of ethnicity. The generalization options follow a hierarchical structure (see Figure 2), where

---

moving up the hierarchy generalizes the information to increase privacy at the cost of utility[55]. For each policy, county, and case number combination, the framework generates 1,000 PK11 estimates. A policy meets the threshold when the upper bound of the estimates' 95% quantile range is less than or equal to 0.01. We choose to evaluate a policy in this manner to increase the likelihood supported policies meet the privacy risk threshold in application. Note, the data sharer can adjust the size of the quantile range to modify the confidence a policy will meet a specific privacy risk threshold.

**Dynamic policy evaluation**
We use the summarized policy search results and forecasted COVID-19 disease case counts to evaluate dynamic policy selection in the context of the COVID-19 pandemic. In this experiment, we measure the proportion of data releases in which the PK11 likely remains below the policy search threshold of 0.01. The dynamic policy is evaluated for two distinct alternative data sharing scenarios: 1) a daily release schedule with a 1-day lagging period assumption and 2) a weekly release schedule. The daily release schedule shares the actual date of diagnosis, prioritizing date granularity at the potential cost of demographic granularity. The weekly release schedule generalizes the date to week of diagnosis.

For each county, the dynamic policy method selects the generalization policy from the search results at the beginning of each week according to the forecasted COVID-19 case volumes. We use the CDC COVID-19 ensemble model's county-specific, one-week forecasts for its superior accuracy over other models[38,56,57]. For the evaluation, we collected all model predictions from August 2020 through April 2021. We obtain daily increase predictions by uniformly distributing the weekly increase point estimate. In selecting policies for the daily release schedule, we use the minimum number of predicted cases in the week. This applies the most privacy preserving policy to all new cases reported in the week. For the weekly release schedule, we use the forecasted one-week increase.

After selecting the sequence of policies for each county, we estimate the privacy risk of sharing the actual reported number of records via the privacy risk estimation framework. We define the actual number of disease cases per day or week by the Johns Hopkins COVID-19 tracking data. The PK11 risk value for each time point in each county is calculated as the upper bound of the 95% quantile range of 1,000 simulations. The evaluation measures the proportion of releases the upper bound remains below 0.01. We additionally evaluate the static application of a policy designed with current, retrospective de-identification techniques, akin to those applied to the CDC's COVID-19 Public Use Data with Geography[11]. The policy, hereafter referred to as the *k*-anonymous policy, shares age intervals in the form (0-17, 18-49, 50-64, and 65+); nearly fully specified race; fully specified ethnicity, sex, and state and county of residence; and date or week of diagnosis. We note the CDC's policy, from which the *k*-anonymous policy derives, was developed to meet regulatory requirements and public health standards under a different release schedule (once every two weeks) and in a retrospective manner (the actual patient records are collected, de-identified and released in a batch). The CDC's policy is designed to achieve 11-anonymity (i.e., PK11 = 0) by generalizing the date of diagnosis to month and by nulling out quasi-identifier information for small groups[11,13,58]. Thus, the *k*-anonymous policy resembles a policy developed with traditional de-identification, but notably differs in its treatment of dates of events and in its assumption of no suppression.

*The complete version of this paper is available, open-access, in the Journal of the American Medical Informatics Association: https://doi.org/10.1093/jamia/ocac011

We further note this last feature is another unique factor to sharing surveillance data in near-real time. Suppression cannot be applied with confidence because it is almost impossible to forecast exactly which records will fall into small demographic groups.

**Case studies**

To provide a specific illustration of the dynamic policy approach to daily releasing updated, record-level disease surveillance data, we consider two Tennessee counties. The first, Davidson County, is a relatively large metropolitan region with a population of approximately 630,000 residents. The second, Perry County, is a relatively rural area with around 8,000 residents.

In each case study, we select a policy on a weekly basis in the same manner as the evaluation. However, to demonstrate how the framework incorporates the data recipient's potential knowledge of diagnosis date, and accounting for the general turnaround time of COVID-19 diagnostic tests results[59–61], we set a 5-day lagging period. Under these constraints, weekly dynamic policy selection first calculates a 5-day rolling sum of new disease case numbers through the coming week. The minimum value of the rolling sum is used to select the policy. We again estimate the privacy risk of sharing the actual number of records under the sequence of selected policies with the privacy risk estimation framework and the Johns Hopkins COVID-19 tracking data. To evaluate the dynamic policy under optimal case load forecasting, we repeat the process by replacing the forecasted case counts with the actual case numbers in policy selection.

**Code**

All experiments are performed using Python (version 3.8). The code, and walkthroughs corresponding to each experiment, can be found at https://github.com/vanderbiltheads/PandemicDataPrivacy

**RESULTS**

**Dynamic policy search**

We summarize the policy search results in Figure 3. To aid in readability, we represent the generalization of each quasi-identifier in a policy with a four-character alphanumeric code. From left to right, the characters represent the age, race, sex, and ethnicity generalizations. We further summarize the results by categorizing US counties by population size.

Once a generalization policy meets the PK11 threshold for a given number of cases, it is unlikely records fall into a demographic group of size 10 or less. Further increasing the case volume increases the number of records in each group and decreases the PK11 value. As such, a policy is listed under the smallest case quantity at which the policy meets the PK11 threshold for every county in the category. It should also be noted there exists a parent-child relationship between policies. For example, policy 2*** is the parent of policy 3***, where the former only differs from the latter by generalizing age to a lesser degree. When a parent policy meets the PK11 threshold, all its child policies also meet the threshold.

As Figure 3 displays, the number of acceptable policies increases with the number of new cases. In most cases, larger counties achieve more acceptable policies than smaller counties at a given case quantity. The maximum number of acceptable policies is

---

73. The most granular policies across all county categories are 1C*e, 2Bse, and 3Ase. Each of these policies prioritizes different types of information. Policy 1C*e offers the most granular age information at the cost of race and sex information, while Policy 3Ase reduces age granularity to increase race and sex specificity.

The case number values are window-size agnostic, such that the policy search results hold regardless of the time period considered. For example, assume a county with fewer than 1,000 residents updates its disease surveillance dataset daily. Further, assume the county adjusts for sets a 5-day lagging period assumption. When the expected number of new cases from the current day and the previous two days sum to 50, the current day's records should be generalized according to either policy **** or **s*. The same policies are supported if, instead, the dataset is updated weekly (and diagnosis date is generalized to week of diagnosis) and 50 new cases are expected for the current week.

**Dynamic policy evaluation**

We summarize the evaluation results, categorizing counties in the same manner as the policy search, in Table 2. There are several major findings. First, dynamically adapting the generalization policy meets the PK11 threshold more frequently than statically applying the $k$-anonymous policy. On average, the dynamic policy meets the threshold for at least 92.8% of the 448 daily releases and 96.0% of the 64 weekly releases. The $k$-anonymous policy meets the threshold as few as 11.8% of the daily releases and 0.4% of the weekly releases. Second, we find that new cases do not occur every day or every week, particularly in counties with fewer residents. As such, there are fewer days the PK11 upper bound can potentially exceed the threshold, inflating proportions in smaller counties.

**Table 2.** Average proportion of time periods where the upper bound of the 95% quantile range of the PK11 risk is less than or equal to 0.01 in the COVID-19 pandemic (August 2, 2020 to October 23, 2021). The average and 95% quantile range in each cell are taken across all counties in the corresponding population size category. The $k$-anonymous policy shares age intervals (0-17, 18-49, 50-64, and 65+), race (Black or African American, White, Asian, American Indian or Alaskan Native, Native Hawaiian or Pacific Islander, Multiple/Other), ethnicity (Hispanic-Latino and Non-Hispanic), sex (Female and Male), and state and county of residency. The $k$-anonymous policy is statically applied to each release. The daily release PK11 estimates apply a 1-day lagging period, while the weekly release estimates assume the actual date of diagnosis is generalized to week of diagnosis.

| County Population Size | Average proportion of daily releases that meet the PK11 threshold in the COVID-19 pandemic [95% Quantile Range] ($n$ = 448) | | Average proportion of weekly releases that meet the PK11 threshold in the COVID-19 pandemic [95% Quantile Range] ($n$ = 64) | |
|---|---|---|---|---|
| | $k$-anonymous Policy | Dynamic Policy | $k$-anonymous Policy | Dynamic Policy |
| < 1,000 ($n$ = 35) | 0.900 [0.790, 0.998] | 1 [1, 1] | 0.605 [0.266, 0.987] | 0.999 [0.984, 1] |

*The complete version of this paper is available, open-access, in the Journal of the American Medical Informatics Association: https://doi.org/10.1093/jamia/ocac011

| | | | | |
|---|---|---|---|---|
| 1,000 - 50,000 (n = 2,129) | 0.389 [0.118, 0.815] | 0.971 [0.902, 1] | 0.072 [0, 0.406] | 0.960 [0.906, 1] |
| 50,000 - 100,000 (n = 398) | 0.181 [0.042, 0.532] | 0.928 [0.868, 0.987] | 0.004 [0, 0.031] | 0.974 [0.922, 1] |
| 100,000 - 1,000,000 (n = 538) | 0.145 [0.009, 0.521] | 0.947 [0.882, 0.998] | 0.008 [0, 0.026] | 0.982 [0.938, 1] |
| > 1,000,000 (n = 39) | 0.118 [0.007, 0.304] | 0.961 [0.874, 0.998] | 0.057 [0, 0.288] | 0.962 [0.906, 1] |

**Case Study: Davidson County, TN**

Figure 4 shows how the forecasted case volumes do not match the weekly seasonality of the actual reported cases in Davidson County. Consequently, the CDC ensemble model tends to overestimate case loads, leading to the selection of more granular policies. Despite the rippling effects of the overestimation, the 95% quantile range of the forecast-driven PK11 remains below 0.01 throughout most of the time frame. Several days exceed the threshold, most of which occur when the selected policies disagree whether to share record-level data under the **** policy or to not share. When sharing fewer than 11 new case records in a 5-day window under the forecast-driven dynamic policy, all new records fall into a demographic group smaller than size 11, resulting in a PK11 of 1.0. Notably, the PK11 never exceeds the threshold when selecting policies according to the actual case counts. Adapting the policy according to perfect forecasts provides optimal privacy protection.

**Case Study: Perry County, TN**

Figure 5 shows that case counts remain relatively small before, as well as after, infection spikes in October 2020 and August 2021. Throughout most of these intervals of low-infection rates, the selected policies from each data source indicate that record-level data should not be shared on a daily basis. However, when the 5-day rolling sums oscillate around 11 cases, the forecasted values again overestimate the weekly minimum case loads, resulting in a PK11 of 1.0. Despite the privacy leaks in the forecast-driven dynamic policy, the dynamic policy guided by the actual disease case counts again maintains the PK11 values below the threshold throughout the time frame.

**DISCUSSION**

This paper introduces a framework to dynamically adjust data sharing policies to publicly share infectious disease surveillance data. The framework forecasts privacy risk according to the expected volume of new cases, enabling data sharers to prospectively adapt policies before seeing case loads. We demonstrate how dynamically changing the policy per the framework's recommendations maintains the privacy risk below the specified privacy risk threshold more frequently than statically applying a policy developed through retrospective de-identification methods, for both the PK and marketer risk-based approaches. The dynamic policy also enhances surveillance utility by fluctuating data generalization with the infection rate, allowing the data sharer to prioritize sharing certain patient information; bypassing the delay of accumulating patient records before performing a risk assessment; and sharing dates of events. These last two features are crucial for characterizing disease transmission[2,31]. Forecasting also enables greater consistency in quasi-identifier representation, as the policy can be maintained throughout the forecasted interval of time. Moreover, predicting which policies provide sufficient privacy protection could potentially automate patient de-identification.

---

*The complete version of this paper is available, open-access, in the Journal of the American Medical Informatics Association: https://doi.org/10.1093/jamia/ocac011

We demonstrate two approaches to dynamic policy adaptation. In the PK risk-based approach, we fix county of residence and date of diagnosis granularity while varying the demographic granularity. We make this tradeoff to support consistent data updates but acknowledge that it may induce certain data utility constraints. For instance, if an application requires uniform demographic granularity, the demographic values may need to be further generalized. An alternative dynamic policy approach could preserve the demographic granularity over time by using the privacy risk estimation framework's predictions to generalize the date of diagnosis into variably-sized time windows. Still, this would impose a utility constraint on date information and cause the data publication schedule to vary. In the marketer risk-based approach (see the Supplementary Information), we show that when the potential attacker has less background knowledge, the dynamic policy can preserve date of diagnosis granularity while monotonically increasing the demographic granularity of the entire dataset over time.

We do not advocate for which measure provides the best privacy protection, nor do we specify which applications each approach best supports; rather, this investigation shows how the privacy risk estimation framework's flexibility can inform different approaches to dynamic policy adjustment.

Despite the merits of this work, we wish to highlight several limitations to guide future extensions and transition into application. First, the dynamic, forecast-driven approach did not always meet the privacy risk threshold in the PK risk-based scenario. However, the framework's policy search results remain relatively robust. Policies chosen from forecasted counts are typically similar or close to those chosen from actual case counts. And when overestimating the number of cases, the privacy risk does not always dramatically exceed the threshold. Furthermore, we selected policies according to a 95% empirical confidence interval, but the policy search can readily incorporate larger confidence intervals as organizations deem desirable. Expanding the intervals further increases the likelihood the dynamic policy will meet the threshold in application. Moreover, when adjusting policies according to the actual case counts, the privacy risk never exceeds the threshold. Thus, the dynamic policy approach can be improved through more accurate forecasts and a model that accounts for potential case load overestimation.

Second, our approach does not incorporate suppression to protect the most unique patient records in the dataset. This is because it is nearly impossible to accurately forecast the exact records which will fall into small demographic groups. It is possible, however, during the enforcement of a selected policy (using the framework) to suppress actual patient records that need to be published and fall into population demographic bins corresponding to very few individuals, such as patient records that are population uniques, or patient records that correspond to population groups with fewer than *k* individuals (for PK risk). Such records with certainty would not meet the *k*-anonymity requirement.

Additional risk analysis can be performed to estimate the risk of actual records in not meeting the *k*-anonymity requirement in a data release and suppress fields in records that are associated with a high estimated risk. Still, the framework's policy search and the policy selection approach depend on many adjustable parameters (e.g., the number of performed simulations, the expected number of new disease cases, the specific bins randomly selected to simulate new cases, the size of the quantile range used for the confidence a policy will meet a given risk threshold), which can be adjusted to mitigate the need for suppression.

Third, as we aim to generally support public data sharing, we focus on privacy risk without measuring the utility of a data generalization policy. Though we provide the data sharer with policy options, from which they can choose how to prioritize sharing quasi-identifier information, and our approach generally supports surveillance utility in terms of providing granular date information and timely updates, we do not address the more complex problem of policy planning. For instance, maximizing the granularity of one quasi-identifier early in the time series could hinder policy flexibility in the future. In the scenario where another quasi-identifier becomes important to public health research later, the data sharer may want to change the generalization of previously released data to complement the new priority. However, if the earlier policy has already consumed the available privacy risk, the policy may not be altered without potentially exposing patients' identities. Previously released data may be shared again with more detail, but not less. Future work should quantitatively measure data utility to inform data sharers in policy planning.

Fourth, the privacy risk estimation framework depends on random sampling methods that may not realistically simulate the pandemic spread of disease. We assign an equal likelihood of infection to all uninfected county residents at any given time in the simulations, and do not allow reinfections. In reality, the actual likelihood varies according to contact patterns of infectious individuals (i.e., through households or at work)[62,63], and reinfections are possible, though not likely in the case of COVID-19[40]. Still, we believe that Monte Carlo simulations, constrained to run within the relatively contained geographic region of a county, provide a reasonable range and estimate of infection outcomes, as they have shown to be adept at simulating complex, high-dimensional patterns[64]. Further framework refinement should address the possibility of reinfection for diseases for which reinfection is more likely.

Fifth, the framework does not compute the re-identification risk of sharing a specific record. Rather, it estimates the range and expectation of privacy risk for a population. Future work should evaluate how well the framework's estimates compare to the re-identification risk of sharing actual disease surveillance data.

Finally, while this paper focuses on de-identification through generalization, an alternative approach would rely on the principle of differential privacy. Differential privacy offers formal privacy guarantees[65]; but as has been recently noted[66], realizing this definition in practice requires injecting noise into the data, a strategy that is not appropriate for every data sharing scenario. Moreover, the CDC's COVID-19 datasets apply generalization and suppression[13]. Therefore, to be consistent with the CDC's current practice, we focused our framework's application on data generalization policies.

**CONCLUSION**

Disease surveillance data is variable, between geographic areas and over time. As such, data must be consistently updated in a timely manner. To support public health research and the public's situational awareness during a pandemic, the data must also contain granular date information. The privacy risk estimation framework we propose enables a prospective approach to surveillance data de-identification. In contrast to traditional methods, prospective policy selection offers increased flexibility, with intermittent consistency, to support near-real time data dissemination. Moreover, we show that forecast-driven de-identification offers better privacy protection than the static data sharing policy application.

**ACKNOWLEDGMENTS**

The authors would like to acknowledge their funding sources: grants 2029651 and 2029661 from the National Science Foundation and training grant T15LM007450 from the National Library of Medicine.

*The complete version of this paper is available, open-access, in the Journal of the American Medical Informatics Association: https://doi.org/10.1093/jamia/ocac011

## AUTHOR CONTRIBUTIONS

J.T.B designed the framework and privacy model, wrote the computer code, performed the experiments, analyzed the results, and prepared the manuscript. C.Y., W.X, Z.Y., and Z.W. contributed to the conceptual design of the framework and privacy model, analyzed the results, and revised the manuscript. A.G.-D., M.K., and M.A.B supervised each component of the project.

**COMPETING INTERESTS**

The authors have no competing interests to disclose.

**AVAILABILITY OF DATA AND MATERIAL**

All data used herein are publicly available. The datasets include: the United States Census PCT12 Tables[37], the Johns Hopkins COVID-19 tracking data[39], and the CDC COVID-19 Ensemble Forecasts[38,67].

# REFERENCES

1. Ibrahim NK. Epidemiologic surveillance for controlling Covid-19 pandemic: types, challenges and implications. J Infect Public Health. 2020 Nov;13(11):1630–8.
2. Thacker SB, Qualters JR, Lee LM. Public Health Surveillance in the United States: Evolution and Challenges* [Internet]. Available from: https://www.cdc.gov/MMWR/preview/mmwrhtml/su6103a2.htm
3. Bansal S, Chowell G, Simonsen L, Vespignani A, Viboud C. Big Data for Infectious Disease Surveillance and Modeling. The Journal of Infectious Diseases. 2016 Dec 1;214(suppl_4):S375–9.
4. Rivers C, Chretien J-P, Riley S, Pavlin JA, Woodward A, Brett-Major D, et al. Using "outbreak science" to strengthen the use of models during epidemics. Nature Communications. 2019 Jul 15;10(1):3102.
5. Woolhouse MEJ, Rambaut A, Kellam P. Lessons from Ebola: Improving infectious disease surveillance to inform outbreak management. Science Translational Medicine. 2015 Sep 30;7(307):307rv5-307rv5.
6. Fang Y, Nie Y, Penny M. Transmission dynamics of the COVID-19 outbreak and effectiveness of government interventions: A data-driven analysis. J Med Virol [Internet]. 2020 Mar 16; Available from: https://www.ncbi.nlm.nih.gov/pmc/articles/PMC7228381/
7. Maybank A. Why racial and ethnic data on COVID-19's impact is badly needed [Internet]. American Medical Association. 2020. Available from: https://www.ama-assn.org/about/leadership/why-racial-and-ethnic-data-covid-19-s-impact-badly-needed
8. Executive Order on Ensuring a Data-Driven Response to COVID-19 and Future High-Consequence Public Health Threats [Internet]. The White House. 2021. Available from: https://www.whitehouse.gov/briefing-room/presidential-actions/2021/01/21/executive-order-ensuring-a-data-driven-response-to-covid-19-and-future-high-consequence-public-health-threats/
9. Haendel MA, Chute CG, Bennett TD, Eichmann DA, Guinney J, Kibbe WA, et al. The National COVID Cohort Collaborative (N3C): Rationale, design, infrastructure, and deployment. Journal of the American Medical Informatics Association [Internet]. 2020 Aug 17;(ocaa196). Available from: https://doi.org/10.1093/jamia/ocaa196
10. Datavant. COVID-19 Research Database [Internet]. Available from: https://covid19researchdatabase.org/
11. COVID-19 Case Surveillance Public Use Data with Geography | Data | Centers for Disease Control and Prevention [Internet]. Available from: https://data.cdc.gov/Case-Surveillance/COVID-19-Case-Surveillance-Public-Use-Data-with-Ge/n8mc-b4w4
12. COVID-19 Case Surveillance Restricted Access Detailed Data | Data | Centers for Disease Control and Prevention [Internet]. Available from: https://data.cdc.gov/Case-Surveillance/COVID-19-Case-Surveillance-Restricted-Access-Detai/mbd7-r32t
13. Lee B, Dupervil B, Deputy NP, Duck W, Soroka S, Bottichio L, et al. Protecting Privacy and Transforming COVID-19 Case Surveillance Datasets for Public Use. Public Health Rep. 2021 Jun 17;00333549211026817.
14. Maxmen A. Massive Google-funded COVID database will track variants and immunity. Nature [Internet]. 2021 Feb 24; Available from: https://www.nature.com/articles/d41586-021-00490-5

15. Gardner L, Ratcliff J, Dong E, Katz A. A need for open public data standards and sharing in light of COVID-19. The Lancet Infectious Diseases [Internet]. 2020 Aug 10;0(0). Available from: https://www.thelancet.com/journals/laninf/article/PIIS1473-3099(20)30635-6/abstract
16. Standards for Privacy of Individually Identifiable Health Information [Internet]. Federal Register. 2000. Available from: https://www.federalregister.gov/documents/2000/12/28/00-32678/standards-for-privacy-of-individually-identifiable-health-information
17. California Consumer Privacy Act (CCPA) [Internet]. State of California - Department of Justice - Office of the Attorney General. 2018. Available from: https://oag.ca.gov/privacy/ccpa
18. Virginia Consumer Data Protection Act Signed Into Law | Lerman Senter [Internet]. Available from: https://www.lermansenter.com/internet-e-commerce/2021/03/08/virginia-consumer-data-protection-act/
19. And Now There are Three …. The Colorado Privacy Act [Internet]. The National Law Review. Available from: https://www.natlawreview.com/article/and-now-there-are-three-colorado-privacy-act
20. Golle P. Revisiting the uniqueness of simple demographics in the US population. In: Proceedings of the 5th ACM workshop on Privacy in electronic society [Internet]. New York, NY, USA: Association for Computing Machinery; 2006. p. 77–80. (WPES '06). Available from: http://doi.org/10.1145/1179601.1179615
21. Rocher L, Hendrickx JM, de Montjoye Y-A. Estimating the success of re-identifications in incomplete datasets using generative models. Nature Communications. 2019 Jul 23;10(1):3069.
22. Sweeney L. Simple Demographics Often Identify People Uniquely. Carnegie Mellon University, Data Privacy Working Paper 3. 2000;34.
23. Benitez K, Malin B. Evaluating re-identification risks with respect to the HIPAA privacy rule. J Am Med Inform Assoc. 2010 Mar 1;17(2):169–77.
24. El Emam K, Dankar FK. Protecting Privacy Using k-Anonymity. J Am Med Inform Assoc. 2008;15(5):627–37.
25. Ohm P. Broken Promises of Privacy: Responding to the Surprising Failure of Anonymization. UCLA L Rev. 2009 2010;57(6):1701–78.
26. Piller C. Data secrecy may cripple U.S. attempts to slow pandemic. Science. 2020 Jul 24;369(6502):356–8.
27. Maxmen A. Why the United States is having a coronavirus data crisis. Nature [Internet]. 2020 Aug 25; Available from: https://www.nature.com/articles/d41586-020-02478-z
28. Office for Civil Rights (OCR). Methods for De-identification of PHI [Internet]. HHS.gov. 2012. Available from: https://www.hhs.gov/hipaa/for-professionals/privacy/special-topics/de-identification/index.html
29. Cassa CA, Grannis SJ, Overhage JM, Mandl KD. A Context-sensitive Approach to Anonymizing Spatial Surveillance Data: Impact on Outbreak Detection. Journal of the American Medical Informatics Association. 2006 Mar 1;13(2):160–5.
30. Gkoulalas-Divanis A, Loukides G, Sun J. Publishing data from electronic health records while preserving privacy: A survey of algorithms. Journal of Biomedical Informatics. 2014 Aug 1;50:4–19.

31. Hope K, Durrheim DN, d'Espaignet ET, Dalton C. Syndromic surveillance: is it a useful tool for local outbreak detection? J Epidemiol Community Health. 2006 May;60(5):374–5.
32. Sun K, Chen J, Viboud C. Early epidemiological analysis of the coronavirus disease 2019 outbreak based on crowdsourced data: a population-level observational study. Lancet Digit Health. 2020 Apr;2(4):e201–8.
33. Malin B, Sweeney L. How (not) to protect genomic data privacy in a distributed network: using trail re-identification to evaluate and design anonymity protection systems. J Biomed Inform. 2004 Jun;37(3):179–92.
34. Samreth D, Arnavielhe S, Ingenrieth F, Bedbrook A, Onorato GL, Murray R, et al. Geolocation with respect to personal privacy for the Allergy Diary app - a MASK study. World Allergy Organization Journal. 2018 Jul 16;11(1):15.
35. Hauser C. Is Your Vaccine Card Selfie a Gift for Scammers? Maybe. The New York Times [Internet]. 2021 Feb 6; Available from: https://www.nytimes.com/2021/02/06/health/covid-vaccination-card.html
36. Kempe A, Beaty BL, Steiner JF, Pearson KA, Lowery NE, Daley MF, et al. The Regional Immunization Registry as a Public Health Tool for Improving Clinical Practice and Guiding Immunization Delivery Policy. Am J Public Health. 2004 Jun;94(6):967–72.
37. Population Census Tables [Internet]. The United States Census Bureau. 2016. Available from: https://www.census.gov/data/datasets/2010/dec/summary-file-1.html
38. Ray EL, Wattanachit N, Niemi J, Kanji AH, House K, Cramer EY, et al. Ensemble Forecasts of Coronavirus Disease 2019 (COVID-19) in the U.S. medRxiv. 2020 Aug 22;2020.08.19.20177493.
39. Dong E, Du H, Gardner L. An interactive web-based dashboard to track COVID-19 in real time. The Lancet Infectious Diseases. 2020 May 1;20(5):533–4.
40. Hall V, Foulkes S, Charlett A, Atti A, Monk EJM, Simmons R, et al. Do antibody positive healthcare workers have lower SARS-CoV-2 infection rates than antibody negative healthcare workers? Large multi-centre prospective cohort study (the SIREN study), England: June to November 2020. medRxiv. 2021 Jan 15;2021.01.13.21249642.
41. Walters CE, Meslé MMI, Hall IM. Modelling the global spread of diseases: A review of current practice and capability. Epidemics. 2018 Dec;25:1–8.
42. Skinner CJ, Holmes DJ. Estimating the Re-identification Risk Per Record in Microdata. Journal of Official Statistics. 1998 Dec;14(4):361.
43. Skinner CJ, Elliot MJ. A Measure of Disclosure Risk for Microdata. Journal of the Royal Statistical Society Series B (Statistical Methodology). 2002;64(4):855–67.
44. CMS Cell Size Suppression Policy | ResDAC [Internet]. Available from: https://www.resdac.org/articles/cms-cell-size-suppression-policy
45. California Department of Health Data De-identification Guidelines (DDG) [Internet]. [cited 2021 Jan 25]. Available from: https://www.dhcs.ca.gov/dataandstats/Documents/DHCS-DDG-V2.0-120116.pdf
46. Utah Department of Health Data Suppression/Data Aggregation Guidelines Summary [Internet]. [cited 2021 Jan 25]. Available from: https://ibis.health.utah.gov/ibisph-view/pdf/resource/DataSuppressionSummary.pdf
47. Washington Department of Health Agency Standards for Reporting Data with Small Numbers [Internet]. [cited 2021 Jan 25]. Available from: https://www.doh.wa.gov/Portals/1/Documents/1500/SmallNumbers.pdf

48. Missouri Department of Health. Data Release Policy | HIV/AIDS Disease Surveillance | Health & Senior Services [Internet]. Available from: https://health.mo.gov/data/hivstdaids/datareleasepolicy.php
49. Sanyaolu A, Okorie C, Marinkovic A, Patidar R, Younis K, Desai P, et al. Comorbidity and its Impact on Patients with COVID-19. SN Compr Clin Med. 2020 Jun 25;1–8.
50. Loukides G, Denny JC, Malin B. The disclosure of diagnosis codes can breach research participants' privacy. J Am Med Inform Assoc. 2010;17(3):322–7.
51. Lee B, Dupervil B, Deputy NP, Duck W, Soroka S, Bottichio L, et al. Protecting Privacy and Transforming COVID-19 Case Surveillance Datasets for Public Use. arXiv:210105093 [cs] [Internet]. 2021 Jan 13 [cited 2021 May 31]; Available from: http://arxiv.org/abs/2101.05093
52. Barth-Jones D. The "Re-Identification" of Governor William Weld's Medical Information: A Critical Re-Examination of Health Data Identification Risks and Privacy Protections, Then and Now [Internet]. Rochester, NY: Social Science Research Network; 2012 Jul. Report No.: ID 2076397. Available from: https://papers.ssrn.com/abstract=2076397
53. Dankar FK, El Emam K. A method for evaluating marketer re-identification risk. In: Proceedings of the 1st International Workshop on Data Semantics - DataSem '10 [Internet]. Lausanne, Switzerland: ACM Press; 2010. p. 1. Available from: http://portal.acm.org/citation.cfm?doid=1754239.1754271
54. Xia W, Liu Y, Wan Z, Vorobeychik Y, Kantacioglu M, Nyemba S, et al. Enabling realistic health data re-identification risk assessment through adversarial modeling. Journal of the American Medical Informatics Association [Internet]. 2021 Jan 15;(ocaa327). Available from: https://doi.org/10.1093/jamia/ocaa327
55. Sweeney L. ACHIEVING k-ANONYMITY PRIVACY PROTECTION USING GENERALIZATION AND SUPPRESSION. Int J Unc Fuzz Knowl Based Syst. 2002 Oct;10(05):571–88.
56. Ray EL, Reich NG. Prediction of infectious disease epidemics via weighted density ensembles. PLOS Computational Biology. 2018 Feb 20;14(2):e1005910.
57. Reich NG, McGowan CJ, Yamana TK, Tushar A, Ray EL, Osthus D, et al. Accuracy of real-time multi-model ensemble forecasts for seasonal influenza in the U.S. PLOS Computational Biology. 2019 Nov 22;15(11):e1007486.
58. Samarati P, Sweeney L. Protecting Privacy when Disclosing Information: k-Anonymity and Its Enforcement through Generalization and Suppression. In: Proceedings of the IEEE Symposium on Research in Security and Privacy (S&P). Oakland, CA; 1998.
59. Tennessee Department of Health. TDH Announces Testing Schedule Change [Internet]. Available from: https://www.tn.gov/health/news/2020/12/14/tdh-announces-testing-schedule-change.html
60. Virginia Department of Health. COVID-19 FAQ [Internet]. Available from: https://www.vdh.virginia.gov/covid-19-faq/, https://www.vdh.virginia.gov/covid-19-faq/
61. County of Los Angeles. COVID-19: Frequently asked questions about testing [Internet]. COUNTY OF LOS ANGELES. 2020. Available from: https://covid19.lacounty.gov/testing-faq/
62. Xie G. A novel Monte Carlo simulation procedure for modelling COVID-19 spread over time. Scientific Reports. 2020 Aug 4;10(1):13120.
63. Schneider KA, Ngwa GA, Schwehm M, Eichner L, Eichner M. The COVID-19 pandemic preparedness simulation tool: CovidSIM. BMC Infectious Diseases. 2020 Nov 19;20(1):859.
64. Metropolis N, Ulam S. The Monte Carlo Method. Journal of the American Statistical Association. 1949 Sep 1;44(247):335–41.

65. Dwork C. Differential privacy. In: Proceedings of the 33rd international conference on Automata, Languages and Programming - Volume Part II [Internet]. Berlin, Heidelberg: Springer-Verlag; 2006. p. 1–12. (ICALP'06). Available from: https://doi.org/10.1007/11787006_1
66. Domingo-Ferrer J, Sanchez D, Blanco-Justicia A. The Limits of Differential Privacy (and Its Misuse in Data Release and Machine Learning) [Internet]. Available from: https://cacm.acm.org/magazines/2021/7/253460-the-limits-of-differential-privacy-and-its-misuse-in-data-release-and-machine-learning/fulltext
67. Center for Disease Control and Prevention. Coronavirus Disease 2019 (COVID-19) [Internet]. Centers for Disease Control and Prevention. 2020. Available from: https://www.cdc.gov/coronavirus/2019-ncov/science/forecasting/forecasts-cases.html
68. Wan Z, Vorobeychik Y, Xia W, Clayton EW, Kantarcioglu M, Ganta R, et al. A Game Theoretic Framework for Analyzing Re-Identification Risk. PLoS One [Internet]. 2015 Mar 25;10(3). Available from: https://www.ncbi.nlm.nih.gov/pmc/articles/PMC4373733/

**FIGURES**
**Figure 1.** Privacy risk estimation framework. The curved rectangles represent processes, the cylinders represent data, and the hexagons represent user-defined parameters. The algorithm that performs the processes within the black box is in the core of the proposed framework, employs Monte Carlo random sampling, and is presented in greater detail in the Methods section. To obtain the privacy risk distributions, the simulation is repeated *n* times. The circled numbers denote the framework steps.

**Figure 2.** The generalization hierarchies for age, race, sex, and ethnicity used in this paper, adapted from those of Wan et al[68]. Each horizontal level is a potential generalization state for the data generalization policy. For example, the policy could specify generalizing age to 5-year age intervals to 15-year age intervals, or broader ranges. We represent year of birth as 1-year age at the bottom of the Age hierarchy. Moving up the hierarchies, the data becomes more generalized to increase privacy. An asterisk indicates the feature is generalized to a null value for all individuals, which is equivalent to suppression or non-release of the corresponding field.

**Figure 3.** Generalization policies with a PK11 upper bound (calculated as the upper bound of the 95% quantile range of 1,000 framework simulations) less than or equal to 0.01 at varying disease case volume thresholds. A four-character alphanumeric code indicates the policy's generalization levels. All policies additionally include state and county of residence and some generalization of diagnosis date. A policy is eligible to be listed under the minimum number of new cases (table column) at which it meets the PK11 threshold for every county in the category (table row). A maximum of two policies are listed in each cell among the actual number of policies supported. The number in the bottom right-hand corner of each cell indicates how many of the 96 searched policies meet the risk threshold at the case volume.

**Figure 4.** Dynamic policy selection applied to Davidson County, TN in the COVID-19 pandemic (August 2, 2020 to October 23, 2021). (Top) The 5-day rolling sum of the forecasted and actual case counts reported in Davidson County. The forecasted counts are from the CDC's COVID-19 ensemble model and the actual counts are from the Johns Hopkins surveillance data. The blue triangles and red squares denote the minimum value within each week (defined as Sunday-Saturday per the CDC model's definition). The minimum values are used to select a policy from policy search results. (Middle) The selected policy at the beginning of each week in the pandemic. Each policy is represented by a 4-character alphanumeric code following the key in Figure 3. The policies are ordered by increasing case count thresholds from bottom to top. Green circles indicate agreement between the policies selected from the forecasted and actual case counts. (Bottom) The PK11 from sharing the actual number of records under the two sequences of policies detailed in the middle graph. The expectation and 95% quantile range are calculated from 1,000 independent framework simulations, while applying a 5-day lagging period assumption. The horizontal dashed line marks the PK11 threshold of 0.01.

**Figure 5.** Dynamic policy selection applied to Perry County, TN in the COVID-19 pandemic (August 2, 2020 to October 23, 2021). (Top) The 5-day rolling sum of the forecasted and actual case counts reported in Davidson County. The forecasted counts are from the CDC's COVID-19 ensemble model and the actual counts are from the Johns Hopkins surveillance data. The blue triangles and red squares denote the minimum value within each week (defined as Sunday-Saturday per the CDC model's definition). The minimum values are used to select a policy from policy search results. (Middle) The selected policy at the beginning of each week in the pandemic. Each policy is represented by a 4-character alphanumeric code following the key in Figure 3. The policies are ordered by increasing case count thresholds from bottom to top. Green circles indicate agreement between the policies selected from the forecasted and actual case counts. (Bottom) The PK11 from sharing the actual number of records under the two sequences of policies detailed in the middle graph. The expectation and 95% quantile range are calculated from 1,000 independent framework simulations, while applying a 5-day lagging period assumption. The quantile ranges are too narrow to be seen outside the mean. The horizontal dashed line marks the PK11 threshold of 0.01.